\documentclass[11pt,a4paper]{article}
\usepackage{graphicx} 
\usepackage{amsmath} 
\usepackage{amssymb} 
\usepackage{hyperref} 
\usepackage{geometry} 
\usepackage{titlesec} 
\usepackage{cite} 
\usepackage{authblk} 
\usepackage{graphicx}
\usepackage[utf8]{inputenc}
\usepackage{enumitem}

\title{\textbf{Embedded Acoustic Intelligence for Automotive Systems}}
\author[1]{Renjith Rajagopal}
\author[1]{Peter Winzell}
\author[1]{Sladjana Strbac}
\author[1]{Konstantin Lindström}
\author[1]{Petter Hörling}
\author[2]{Faisal Kohestani}
\author[2]{Niloofar Mehrzad}
\affil[1]{Volvo Car Corporation}
\affil[2]{Chalmers University of Technology}

\date{\today}

\begin{document}

\maketitle

\begin{abstract}
Transforming sound insights into actionable streams of data, this abstract leverages findings from degree thesis research to enhance automotive system intelligence, enabling us to address road type~\cite{degree_report}.By extracting and interpreting acoustic signatures from microphones installed within the wheelbase of a car, we focus on classifying road type.Utilizing deep neural networks and feature extraction powered by pre-trained models from the Open AI ecosystem (via Hugging Face~\cite{huggingfacehubdocs}), our approach enables Autonomous Driving and Advanced Driver-Assistance Systems (AD/ADAS) to anticipate road surfaces, support adaptive learning for active road noise cancellation, and generate valuable insights for urban planning.
The results of this study were specifically captured to support a compelling business case for next-generation automotive systems. This forward-looking approach not only promises to redefine passenger comfort and improve vehicle safety, but also paves the way for intelligent, data-driven urban road management, making the future of mobility both achievable and sustainable.
\end{abstract}


\section{Introduction}  \vspace{-1.0em}
\textbf{Background Information} \\
\textbf{Road/Tyre Noise Fundamentals} \\
Road and tyre noise is a fundamental component of the external sound produced by vehicles in motion. This noise primarily arises from the interaction between tyres and the road surface, resulting in a combination of rolling, vibration, and frictional sounds. The characteristics of road/tyre noise are shaped by several key factors:
\begin{itemize}
\item Vehicle speed: Higher speeds generally increase the intensity of tyre-road noise, with noise levels rising by approximately 6 dB for each doubling of speed.
\item Road surface texture: Rougher or more textured surfaces generate more pronounced noise due to increased friction and vibration at the contact patch. Smooth asphalt can reduce noise by 2–4 dB compared to coarse surfaces.
\item Tyre characteristics: The type, tread pattern, condition, and width of the tyre all influence the noise produced. Wider tyres can increase noise by about 0.5–1 dB per centimeter of width.
\end{itemize}
At speeds above 30 km/h, tyre-road interaction becomes the dominant source of vehicle noise, produced through three main mechanisms:
\begin{itemize}
\item Tread impact: Pavement texture excites tyre vibrations, typically in the 50–1000 Hz range.
\item Air pumping: Compression and expulsion of air in the tread grooves, generating noise in the 1–4 kHz range.
\item Helmholtz resonances: Cavity oscillations within tyre voids.
\end{itemize}

Modern vehicles are advancing rapidly in both autonomous driving (AD) and in-cabin acoustic technologies. Deep neural networks (DNNs) are central to AD, enabling perception through Light Detection and Ranging(LiDAR), cameras, and radar for object detection and trajectory prediction. However, dynamic and unpredictable environments present challenges that are also mirrored in the audio domain.

Accurate road type classification is essential for autonomous driving (AD) and advanced driver-assistance systems (ADAS), as it enables vehicles to adapt dynamically to changing surface conditions. Traditional sensor-based methods often struggle under environmental challenges such as poor lighting or weather. In contrast, approaches leveraging tire noise analysis and deep learning achieve robust, real-time road surface recognition. This allows AD/ADAS systems to optimize vehicle control—such as braking, acceleration, and stability—based on the identified road type, reducing risks like skidding or loss of traction. Integrating audio-based road type classification provides valuable redundancy for conventional sensors and supports safer navigation in complex or adverse conditions.

Traditionally, in-vehicle noise management relied on passive sound insulation and dynamic dampers, which increase vehicle weight and struggle to attenuate low-frequency infrasound. The introduction of Active Noise Cancellation (ANC)~\cite{WikipediaActiveNoiseControl} systems, which use microphones and controllers to analyze in-cabin noise and emit inverted sound waves, marked a significant shift. However, conventional ANC is most effective against constant, predictable noises and is limited by real-time noise measurement and analysis. Active Road Noise Cancellation (ARNC)~\cite{arnc} technologies address these limitations by detecting vibrations and road noise in real time using accelerometers and microphones, processing these signals with digital signal processors(DSPs)~\cite{WikipediaDigitalSignalProcessor} to generate anti-noise waveforms within milliseconds. This enables rapid and adaptive cancellation of various road noise types, demonstrating up to a 3dB reduction in in-cabin noise and allowing for reduced reliance on heavy insulation materials.

Despite these advances, current ARNC systems operate primarily reactively, responding to detected noise rather than proactively predicting it before it reaches the cabin. Integration of contextual data—such as road surface type, vehicle speed, and position-based mapping—remains limited, constraining the adaptability and effectiveness of these systems. Volvo Cars’ patented concept~\cite{us20230097755} for context-aware cancellation systems exemplifies the industry’s shift toward predictive noise management, proposing the use of local and remote contextual data to generate spatially targeted anti-noise waveforms.This framework outlines individualized noise mitigation through dynamic adjustment of cancellation signals across vehicle zones, though remains a theoretical blueprint rather than deployed infrastructure. Furthermore, studies and methods on how these systems affect occupants indicates that they need to be able to dynamically adapt on individual basis as well as to the contextual surroundings~\cite{us20240345650A1}.

Road type classification also benefits urban planning by providing data-driven insights into different surface types within city environments. This process enables planners to make informed decisions about infrastructure development, utility routing, and land use optimization. By accurately identifying road types, planners can assess construction feasibility, optimize project costs, and minimize environmental disruption. Additionally, road type classification supports the integration of new technologies—such as autonomous vehicles and smart city sensors—by ensuring that digital models and navigation systems reflect real-world conditions, ultimately leading to safer, more efficient, and sustainable urban spaces.

\textbf{Motivation \& Objective of Study} \\
While conceptual frameworks like Volvo’s recent patent~\cite{us20230097755} envision context-aware noise cancellation that leverages road noise information, practical deployment of such systems hinges on the ability to robustly classify the complex acoustic signatures generated by tyre-road interactions. Existing solutions often depend on pre-recorded noise profiles or static models, which lack adaptability to real-world variations in road surfaces, tyre conditions, and environmental factors. This limitation reduces the effectiveness of noise mitigation, especially as vehicles encounter a wide range of dynamic driving conditions.

To address this gap, our work systematically compares two leading deep neural network architectures for road type classification: Convolutional Neural Networks (CNNs), and transformer-based models such as the Audio Spectrogram Transformer (AST). CNNs offers lightweight, efficient inference suitable for deployment on embedded automotive hardware, while AST provides enhanced accuracy through global attention mechanisms, albeit with higher computational demands.

The rapid expansion of open-source AI platforms like HuggingFace~\cite{huggingfacehubdocs} has made both CNN and transformer-based models readily accessible for research and benchmarking in audio classification tasks. However, there remains a critical need to evaluate these architectures not only for their classification accuracy, but also for their suitability in real-time, resource-constrained automotive environments.

Our study is motivated by the need to bridge the gap between academic advances in audio-based classification and the stringent requirements of automotive-grade implementation. By benchmarking these models on real-world road noise data, we aim to enable the next generation of predictive noise cancellation systems—where robust, low-latency classification serves as a key input for adaptive algorithms that generate proactive anti-noise signals tailored to changing road conditions, while simultaneously informing AD/ADAS system for situational awareness and urban infrastructure planning through road type-aware data streams.

\section{Related Work}
\textbf{Traditional and Machine Learning Approaches to Road Noise Analysis} \\
Early road noise management relied on passive insulation and ARNC, utilizing accelerometers and digital signal processors (DSPs) to generate anti-noise waveforms within 2ms of detection~\cite{arnc}. While effective for steady-state noise, these systems struggled with transient events and lacked adaptability to dynamic road conditions. Statistical models and Gaussian mixture models (GMMs) were later employed for traffic noise prediction, but required extensive feature engineering and faced generalization challenges across diverse environments.

\textbf{Deep Learning for Road Surface and Condition Detection} \\
Recent advances leverage deep neural networks (DNNs) for road-type classification~\cite{lee2023road} demonstrated convolutional recurrent neural networks (CRNNs) for terrain recognition, achieving real-time performance on 13 road types using tire-noise data. Similarly, recurrent neural networks (RNNs) for detecting road surface wetness from tire interaction sounds, achieving robustness across speed and pavement variations~\cite{zhang2024road}. However, these approaches focused on spectral features without optimizing for embedded automotive hardware constraints.

\textbf{Gaps in Current Research} \\
While prior work establishes the efficacy of CNNs and transformers in acoustic tasks, three critical gaps persist:
\begin{itemize}
    \item \textbf{Embedded Feasibility:} Most studies validate models on Graphics Processing Unit(GPUs), neglecting memory (\(<50\)MB) and latency (\(<20\)ms) constraints of automotive systems-on-chip (SoCs).
    \item \textbf{Temporal Consistency:} Reactive systems which that enable RANC lack advance prediction windows (\(>50\)ms) for destructive interference, limiting effectiveness against transient noise.
    \item \textbf{Context Integration:} Volvo’s patent outlines a theoretical framework but provides no implementation details for road-noise classifiers or hybrid quantization techniques.
\end{itemize}

This study addresses these gaps by benchmarking CNN and AST on automotive-grade hardware, while establishing temporal stability metrics for embedded intelligence use cases.
\section{Methodology}
This section details the approach for classifying road types using road-tyre noise signatures. The methodology comprises three main components: \textbf{sensing \& data collection}, \textbf{audio preprocessing}, and \textbf{classification}.

\textbf{Sensing \& data collection}: 
A Volvo EX40 EV was retrofitted with a microphone mounted in the wheelbase, configured
to sample at 44.1 kHz for the collection of road tire noise signals. Data was gathered while
driving over various surfaces-including smooth asphalt, rough asphalt, concrete pavement, Belgian pavement, Vienna pavement, and sections with pipes-at the Volvo Cars Hällered Proving
Ground~\cite{VolvoHällered}. The vehicle was operated at controlled speeds ranging from 40 km/h to 90 km/h,
ensuring that tire noise remained the dominant acoustic source during the tests.

\textbf{Audio Preprocessing}: 
The raw audio data undergoes time-frequency analysis, employing techniques originally developed for speech processing. This step extracts salient features and transforms the continuous signal into an $N$-dimensional feature vector for each audio frame, providing a compact and informative representation of the acoustic environment.

\textbf{Classification}: 
The classification pipeline operates in two phases:
\begin{itemize}
    \item \textbf{Training Phase}:
    During the training phase, both a pre-trained CNN model and the transformer-based Audio Spectrogram Transformer (AST) were adapted for road-type classification using audio recordings. Both models were originally designed for a large number of audio classes, but were modified for the specific task of distinguishing between three road surface types: rough asphalt, smooth asphalt, and others.
    This approach leverages the generalization capabilities of deep neural networks while specializing the models for the specific application of road-type classification based on acoustic signatures. The training process focuses on adapting pre-trained models to a new domain, enabling effective feature extraction and accurate classification of road surfaces using vehicle audio data.
    
    \item \textbf{Classification Phase}: 
    The optimized model is deployed to classify incoming audio frames in real time, enabling prompt and accurate identification of road surfaces during vehicle operation.
\end{itemize}

\begin{figure}[h]
  \centering
  \includegraphics[width=0.8\textwidth]{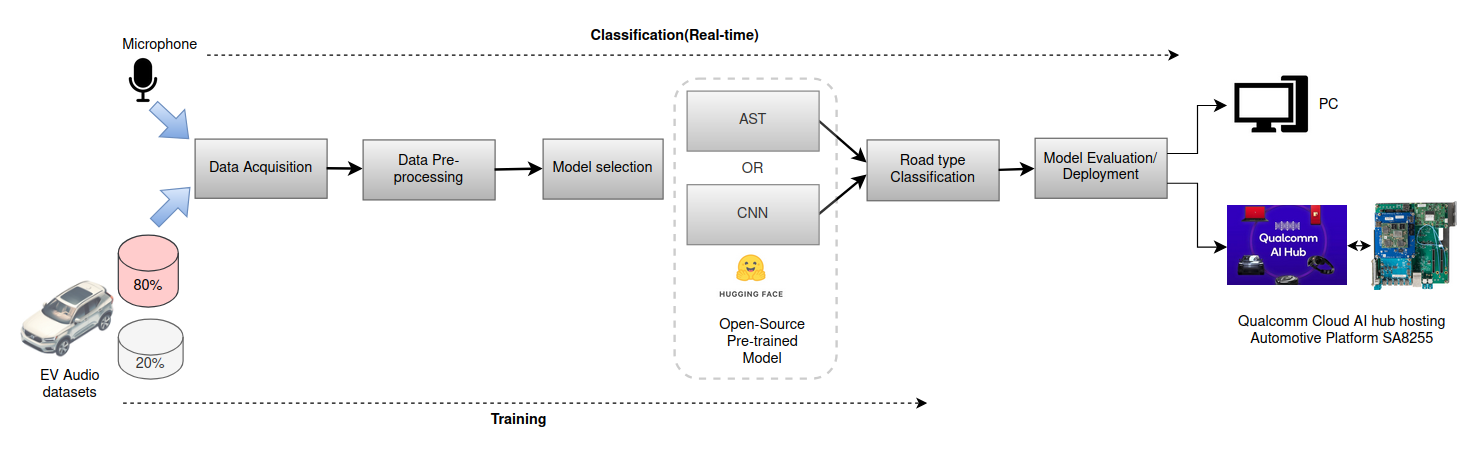}
  \caption{Stages involved in road type classification using road noise signature}
  \label{fig:road-classification}
\end{figure}

Once in the classification stage, the previously trained models were deployed to Embedded Automotive development platform hosted in virtual cloud that interpret and classify the audio captured by the microphone. Figure~\ref{fig:road-classification} illustrates this workflow

\begin{figure}[h]
  \centering
  \includegraphics[width=0.8\textwidth]{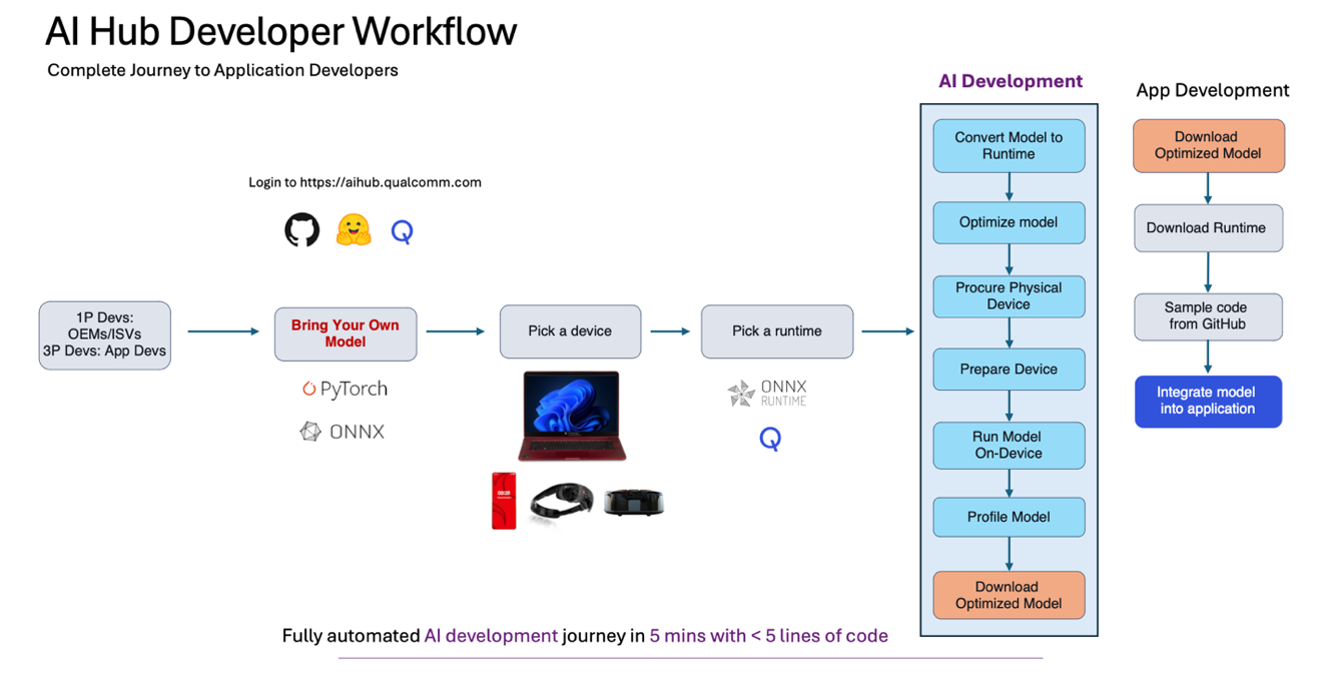}
  \caption{Developer workflow for Model deployment to Automotive Embedded Platform.Reproduced from Qualcomm AI Hub \url{https://app.aihub.qualcomm.com/docs/hub/index.html}}
  \label{fig:model-dep}
\end{figure}

\textbf{Qualcomm AI Hub for Model deployment} \\[0.5em]
The core strength of our approach lies in deploying pre-trained audio classification models on embedded automotive platforms using the Qualcomm AI Hub~\cite{QualcommAIHub}. By leveraging the Automotive Development Platform SA8255~\cite{LantronixRideSX4}, hosted virtually within the AI Hub ecosystem, we can harness the extensive knowledge and robust feature extraction capabilities of models trained on datasets-without the need for extensive on-device training or high computational overhead.Figure~\ref{fig:model-dep} illustrates this workflow

Qualcomm AI Hub enables developers to access and deploy quantization-ready, hardware-optimized models, ensuring efficient and accurate inference on resource-constrained automotive hardware. The platform’s support for model quantization preserves model accuracy and performance while reducing computational demands, making real-time, intelligent in-vehicle audio systems feasible.

A significant advantage of this workflow is the ability to validate, profile, and optimize models on virtual Qualcomm hardware in the cloud, eliminating the need for physical devices during development. This not only reduces development costs and accelerates time-to-market but also allows for rapid iteration and testing.

By capitalizing on the generalization abilities and sophisticated representations of pre-trained models, our strategy opens up new opportunities for deploying advanced machine learning applications in embedded automotive environments, enabling efficient, real-time audio intelligence even with limited computational resources.

Detailed information on the benchmarks of different model architectures (CNN and AST) and techniques used for classification is mentioned in the degree project report made by students from Chalmers University of Technology~\cite{degree_report}.

\textbf{Open-Source AI Systems and Hugging Face} \\[0.5em]
Open-source AI systems have accelerated research and development by making state-of-the-art machine learning models and datasets widely accessible. \textbf{Hugging Face} is a leading open-source platform and model hub that hosts a vast collection of pre-trained models for domains such as natural language processing, computer vision, and audio analysis~\cite{huggingfacehubdocs}. Through resources like the Hugging Face Hub and the Transformers library, researchers and developers can easily find, share, and deploy advanced models---including those for audio classification---without the need to train from scratch. These pre-trained models are designed for rapid adaptation to new tasks and support integration with popular deep learning frameworks. The collaborative ecosystem and extensive educational resources provided by Hugging Face further lower barriers to entry, enabling efficient experimentation, benchmarking, and deployment of AI solutions in both academic and industrial contexts.

\vspace{1em}

\textbf{Combating Open-Washing: Transparent AI for Safety-Critical Automotive Systems} \\[0.5em]
Open-source AI platforms like Hugging Face have made advanced audio classification models---including CNN-based models like YAMNet~\cite{KaggleYAMNet} and transformer-based models like AST---widely accessible for rapid research and deployment. These pre-trained models can be easily adapted and integrated into in-vehicle systems for tasks such as contextual information delivery to AD/ADAS and active road noise control. However, many so-called ``open'' models are only open-weight: they release model parameters but withhold essential components like training code, datasets, or preprocessing pipelines. This practice, known as \textbf{open-washing}, undermines transparency, reproducibility, and fairness, and can introduce regulatory risks---especially for safety-critical automotive applications.

To address these issues, the Linux Foundation has introduced frameworks and tools to define and enforce genuine openness~\cite{LFAIDataMOF2024}.

\begin{itemize}
    \item \textbf{Model Openness Framework (MOF~\cite{isitopenAI}):} Classifies models by openness, from basic open weights (Class III) to full transparency with code, data, and documentation (Class I).
    \item \textbf{Model Openness Tool (MOT):} Assesses and scores the openness of AI models, helping identify open-washing.
    \item \textbf{OpenMDW-1.0 License~\cite{OpenMDWAI}:} A permissive license tailored for AI models, ensuring legal clarity and requiring all critical artifacts to be openly licensed.
    \item \textbf{Open Model Initiative (OMI):} Promotes community-driven, high-quality open models with rigorous transparency standards.
\end{itemize}

For automotive deployments---such as predictive road noise cancellation or AD/ADAS integration---using models validated by MOF/MOT and licensed under OpenMDW-1.0 ensures transparency in data, pipelines, and architecture. This not only supports regulatory compliance and fairness but also fosters innovation and reliability through community collaboration.

\section{Results}
The comparative evaluation of AST and CNN architectures for road-type classification on automotive-grade embedded hardware is systematically presented in Section~4 of the degree report~\cite{degree_report}. Key findings are listed below.

\textbf{Model Performance}

\textbf{CNN (Convolutional Neural Network)} \vspace{-1.0em}
\begin{itemize}
    \setlength{\itemsep}{0pt}
    \setlength{\parskip}{0pt}
    \item Achieved \textbf{highest accuracy} among tested architectures
    \item Demonstrated \textbf{real-time performance} on embedded hardware:
        \begin{itemize}
            \setlength{\itemsep}{0pt}
            \setlength{\parskip}{0pt}
            \item Inference latency consistently \textless 20\,ms
            \item Memory footprint \textless 50\,MB
        \end{itemize}
    \item Suitable for deployment on resource-constrained automotive SoCs
\end{itemize}
\vspace{-1.0em}
\textbf{Transformer (AST)} \vspace{-1.0em}
\begin{itemize}
    \setlength{\itemsep}{0pt}
    \setlength{\parskip}{0pt}
    \item Superior classification accuracy in controlled environments:
        \begin{itemize}
            \setlength{\itemsep}{0pt}
            \setlength{\parskip}{0pt}
            \item Outperformed CNNs on specific datasets
        \end{itemize}
    \item Higher computational requirements:
        \begin{itemize}
            \setlength{\itemsep}{0pt}
            \setlength{\parskip}{0pt}
            \item Increased latency compared to CNNs
            \item Greater memory usage
        \end{itemize}
    \item Optimization improvements:
        \begin{itemize}
            \setlength{\itemsep}{0pt}
            \setlength{\parskip}{0pt}
            \item Quantization reduced resource demands
        \end{itemize}
\end{itemize}
\vspace{-1.0em}
\textbf{Embedded Feasibility}
\vspace{-1.0em}
\begin{itemize}
    \setlength{\itemsep}{0pt}
    \setlength{\parskip}{0pt}
    \item Hardware platform: Automotive reference system with Qualcomm Hexagon Tensor Processor(HTP)~\cite{QualcommAIEngineDirectSDK}
    \item CNN performance:
        \begin{itemize}
            \setlength{\itemsep}{0pt}
            \setlength{\parskip}{0pt}
            \item Met all real-time requirements
            \item Maintained consistent low-latency operation
        \end{itemize}
    \item Transformer considerations:
        \begin{itemize}
            \setlength{\itemsep}{0pt}
            \setlength{\parskip}{0pt}
            \item Required additional optimization for practical deployment
            \item Showed potential for accuracy-sensitive applications
        \end{itemize}
\end{itemize}

\section{Conclusion and Future Works}
The CNN architecture proved most effective for road-type classification using road noise acoustic signatures, offering a balance of accuracy and computational efficiency. Transformer-based models (AST) demonstrated superior classification performance but faced challenges in automotive embedded execution due to higher latency and memory demands, limiting their viability for resource-constrained systems.

This study demonstrates the feasibility of audio-based road type classification using both CNN and transformer architectures, yet significant opportunities remain for future experimentation to address current limitations and advance real-world applicability. The results highlight the importance of dataset quality—particularly in terms of size, class balance, and diversity—which directly impacts model generalization and robustness. Future efforts should prioritize the collection of larger, more balanced datasets, incorporating a broader range of road surfaces, environmental conditions, and vehicle types. This will help mitigate the risk of overfitting and improve model reliability across a wider spectrum of real-world scenarios.

Future experimentation should also focus on integrating uncertainty-aware learning techniques, such as Bayesian inference and probabilistic neural networks, to better handle model overconfidence and improve robustness in open-world settings~\cite{bagaev2023reactive}. Validation in deterministic, hard real-time systems—such as those required for AD/ADAS and adaptive noise cancellation—will be essential to ensure that these models meet the stringent safety and latency requirements of modern automotive applications.

Looking ahead, investigating ISO/PAS 8800~\cite{ISO_PAS_8800_2024}—the emerging standard for AI safety in automotive systems—will be essential to ensure that audio-based road type classification models are developed, validated, and deployed in line with the highest safety and reliability requirements. This proactive approach will help address AI-specific risks and uncertainties, supporting the responsible integration of advanced machine learning into safety-critical vehicle applications.

By addressing these aspects, future experimentation can drive the evolution of audio-based road type classification from a promising research prototype to a reliable, integrated component of next-generation vehicle safety and comfort systems. \\

\bibliographystyle{ieeetr}
\bibliography{references}

\end{document}